\documentclass[prl,aps,twocolumn,superscriptaddress]{revtex4-1}
\usepackage{graphicx,color}
\usepackage{amsthm}
\usepackage{amsfonts}
\usepackage{algorithmic}
\usepackage{enumerate}
\usepackage{latexsym}
\usepackage{amsmath}
\usepackage{amssymb}
\usepackage[colorlinks=true,citecolor=blue,linkcolor=blue]{hyperref}

\usepackage{dcolumn}
\usepackage{bm}

\usepackage[utf8]{inputenc}
\usepackage[T1]{fontenc}
\usepackage{mathptmx}
\newcommand\degree{^\circ}
\newcommand{\tabincell}[2]{\begin{tabular}{@{}#1@{}}#2\end{tabular}}

\emergencystretch=\maxdimen
\begin{document}
\title{Quantum confinement and edge effects on electronic properties of zigzag green phosphorene nanoribbons}

\author{Chi Ma}
\affiliation{Department of Physics, Beijing Normal University, Beijing 100875, China}
\affiliation{College of Integrative Sciences and Arts, Arizona State University, Mesa, Arizona 85212,USA}
\author{Tianxing Ma}
\email{txma@bnu.edu.cn}
\affiliation{Department of Physics, Beijing Normal University, Beijing 100875, China}
\affiliation{Beijing Computational Science Research Center, Beijing 100193, China}

\author{Xihong Peng}
\email{Xihong.peng@asu.edu}
\affiliation{College of Integrative Sciences and Arts, Arizona State University, Mesa, Arizona 85212,USA}

\begin{abstract}
First principles density-functional theory calculations were performed to investigate quantum confinement and edge effects on the electronic properties of zigzag green phosphorene nanoribbons (ZGPNRs) with edge chemical species including H, OH, F, Cl, O, and S for the ribbons width in the range of 0.5 \~{} 3.7 nm. The ZGPNRs were obtained from the relaxed two-dimensional (2D) green phosphorene monolayer with different cutting strategies and the most energetically favorable ribbon configuration was selected for further exploration of the size and edge effects. It was found that the electronic properties of the ZGPNRs are strongly associated with the ribbon width and edge chemical species. They show either semiconducting or metallic features depending on the edge functionalization species. The ZGPNRs show semiconducting behavior with the edge species of H, OH, F, or Cl (Group \uppercase\expandafter{\romannumeral1}), while exhibit metallic characteristics with pristine or O, S edges (Group \uppercase\expandafter{\romannumeral2}). The conduction band minimum (CBM) and valence band maximum (VBM) of the ZGPNRs with the Group \uppercase\expandafter{\romannumeral1} edge are primarily located at the inner P atoms and the edge P and functionalization atoms have little contribution. However, for the Group \uppercase\expandafter{\romannumeral2} edge, the electronic bands crossing the Fermi level are dominantly contributed by the edge atoms. It was also found that the band gap and work function of the ZGPNRs are tunable by varying ribbon width and edge functionalization species.
\end{abstract}


\maketitle
\noindent
\underline{\it Introduction}
Recently, black phosphorene (BP, $\alpha$-phosphorene), a novel two-dimensional (2D) layered semiconductor material, has triggered intensive research interest\cite{RN53,RN39,RN63,RN56,RN16}. Due to its prominent carrier mobility, broad direct band gaps and high in-plane anisotropy, BP has been considered as a promising material for many applications, such as optoelectronics\cite{RN153,RN154}, sensors\cite{RN157,RN158}, batteries\cite{RN155,RN156} and catalysis\cite{RN71,RN143}. Following experimental fabrication of monolayer BP\cite{Castellanos_Gomez_2014} and blue phosphorene ($\beta$-phosphorene)\cite{RN62}, plenty of other allotropes such as$\gamma-, \delta-,\varepsilon-,\zeta-,\eta-,\theta-,\psi-$phosphorene were predicted based on theoretical calculations\cite{RN52,RN57,RN58,RN59,RN60,RN61}. Green phosphorene (GP, $\lambda$-phosphorene) constructed from the combination of blue and black allotropes was initially reported by Han \textit{et. al.} through theoretical calculations\cite{RN21}. Since GP has lower formation energy compared to blue phosphorene and a direct band gap up to 2.4 eV, this material attracts numerous attention in research community, focusing on its mechanical flexibility and strong anisotropy\cite{RN22,RN18,RN20}.

Tailoring material properties is critical for its applications. For phosphorene, many strategies were investigated to engineering its electronic properties such as band gap and those tuning factors include varying its allotropes, number of atomic layers\cite{RN63,RN64}, applying mechanical strain\cite{RN22,RN66}, nano-patterning\cite{RN67}, and field effect\cite{RN24,RN28,RN25,RN65}. Size is also a very commonly used tuning factor to engineer material properties due to quantum confinement effect. Therefore, tailoring 2D sheet into one-dimensional (1D) nanoribbons with different ribbon width attracts particular research interest. Once the 1D ribbons obtained, how to treat the edge dangling bonds will largely affect the properties of the ribbons. For green phosphorene, size and strain effects on mechanical and electronic properties of H-passivated green phosphorene nanoribbons (GPNRs) have been reported by Garrison \textit{et. al.}\cite{RN26}. In this work, we focus on size and edge effects on the zigzag green phosphorene (ZGPNRs) and explore a series of edges including pristine, passivation with H, OH, F, Cl, O, or S for the ribbon width in the range of 0.5 \~{}  3.7 nm. Structural and electronic properties such as band structures, band gap, and work functions were obtained as a function of ribbon size and edge functionalization. Our results suggest that the ZGPNRs show either semiconducting or metallic behavior depending on the edge passivation.
\begin{figure*}
	\includegraphics[width=16cm]{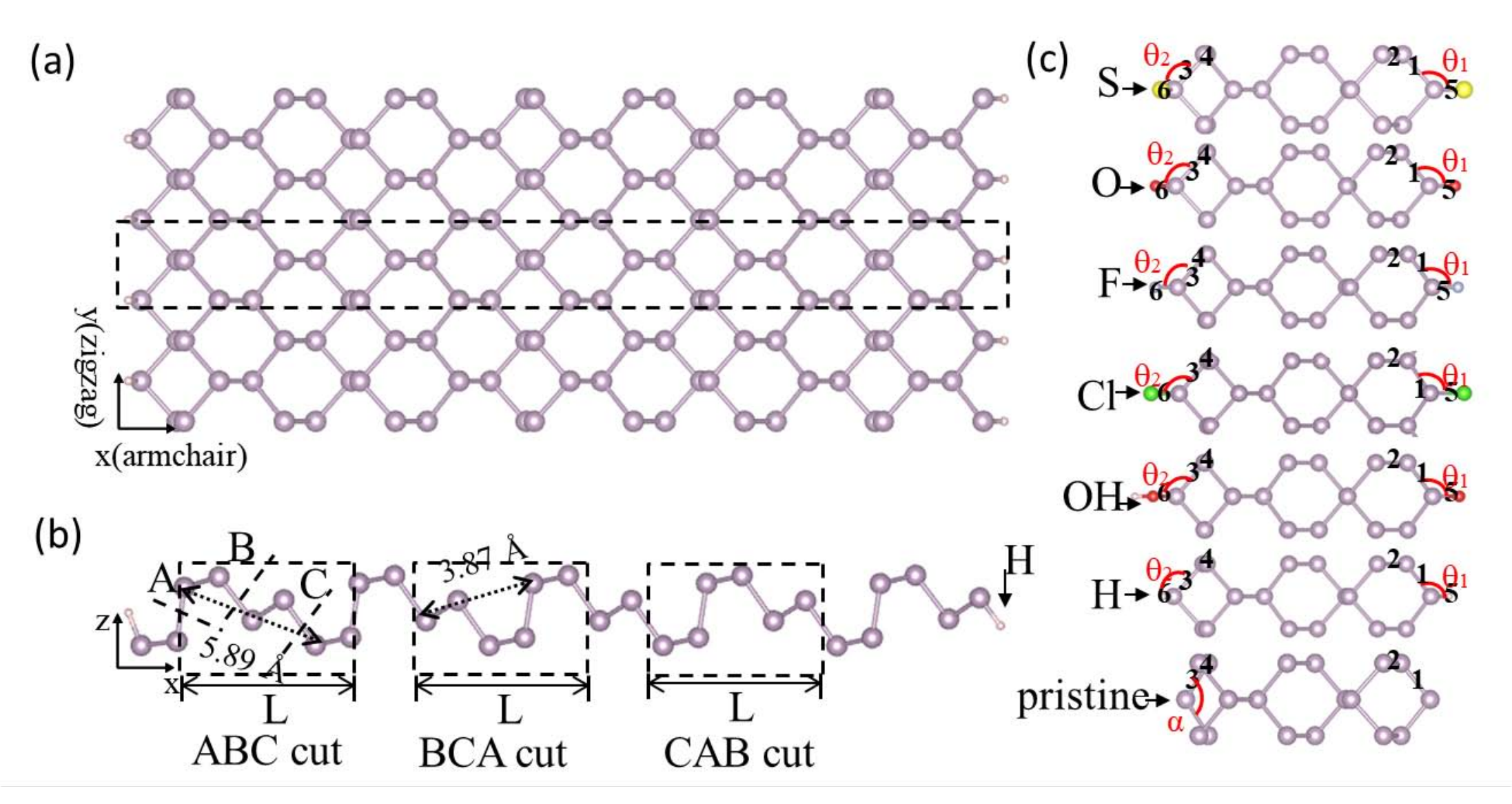}
	\caption{\label{fig1}The structure of green phosphorene nanoribbons   (a)top view; (b)side views; (c) relaxed structures of the pristine 2L-ZGPNRs (i.e. width 14.66 Å) and the ribbons with the edge passivated by H, F, Cl, OH, O, S atoms, respectively}
\end{figure*}

\noindent
\underline{\it Methodology}
The calculations were performed using the first-principles density functional theory (DFT)\cite{RN147} with the Perdew-Burke-Ernzerhof (PBE) exchange-correlation functional\cite{RN148} and projector-augmented wave (PAW) potentials\cite{RN149,RN150}, which were implemented in the Vienna Ab-initio Simulation Package (VASP)\cite{RN151,RN152}. The kinetic energy cut-off for the plane wave basis was set to be 500 eV, and the convergence criteria for the structural optimization are $10^{-5}$ eV and $10^{-4}$ eV for the electronic and ionic iterations, respectively. The Monkhorst-Pack k-point grid of 1$\times$16$\times$1 is utilized for reciprocal space sampling. A total of 21 k-points were sampled in each high symmetry line of the reciprocal space for the electronic band structure calculations. A vacuum space of at least 17 Å was adopted in the unit cell to eliminate the interaction between ribbons resulted from periodic boundary condition. The band gap is defined as the energy difference between the conduction band minimum (CBM) and the valence band maximum (VBM). The work function is calculated by the energy difference between the vacuum and the Fermi level.

\noindent
\underline{\it Result and Discussion:} {\it Crystal structures of ZGPNRs.} The ZGPNRs are truncated along the zigzag direction (i.e. y-axis) from the 2D monolayer GP which was cleaved from bulk green phosphorus. Our calculated lattice constants of bulk green phosphorus in monoclinic C2/m structure are a = b = 7.29 Å and c = 11.50 Å and the relaxed lattice constants of the monolayer green phosphorene are a = b = 7.34 Å, which are in good agreement with other theoretical calculations\cite{RN21,RN22}. GP is a combination of black and blue phosphorene in a ratio of 1:2\cite{RN18}. Compared to black and blue phosphorene, GP has lower symmetry and a lager unit cell. The 2D GP is a trilayer structure as shown in FIG.\ref{fig1}(a) and (b)\cite{RN21,RN22}. To obtain 1D ribbons along the zigzag direction, there are three different ways to trim the 2D sheet, i.e. breaking the A, B, or C bonds as indicated in FIG.\ref{fig1}(b), namely ABC, BCA, and CAB cut, respectively.  The width of the ZGPNRs was referred as the number of the cutting unit marked by the dash line in FIG.\ref{fig1}(b), notated as 1L, 2L, 3L \textit{etc}, where L is the horizontal length of the cutting unit 7.28 Å. For instance, 2L ABC-cut ribbons has ABCABC sequence with width 14.56 Å and 3L BCA ribbons is BCABCABCA with width 21.84 Å. As shown in FIG.\ref{fig1}(b), BCA and CAB-cuts correspond to the same structure because flipping over one you can get the other one. The energies of the pristine ribbons with different cutting sequence and width were calculated and compared in Table \ref{tab:table1}. It shows that the energy of the CAB sequence is slightly lower than that of ABC indicating that the CAB sequence is energetically more stable. The lower energy of the CAB sequence may be resulted from more compacted atomic structure by comparing the fourth-nearest P-P bond distance where the distance is 3.87 Å and 5.89 Å  for the CAB(BCA) and ABC-cut, respectively as shown in FIG.\ref{fig1}(b). The energies of these two different sequences with other edge passivation atoms such as H, OH were also compared and found out that the CAB sequence  has lower energy than ABC regardless the edge configuration. Therefore, the ribbons studied in this work were using the CAB-cut sequence.

\begin{table}
	\caption{\label{tab:table1}total energy  of the pristine ZGPNRs with the ABC and CAB-cutting sequence. }
	\begin{ruledtabular}
		\begin{tabular}{cccccc}
			Ribbon notation (nL)&1L&2L&3L&4L&5L\\
			\hline
			Ribbon width (Å) & 7.28 & 14.56 & 21.84 &29.12&36.40\\
			\hline
			Tol. E. of ABC cut (eV) & -30.38 & -62.57 & -94.75&-126.94&-159.12\\
			Tol. E. of CAB cut (eV) & -30.89 & -62.99 & -95.18&-127.36&-159.53\\
		\end{tabular}
	\end{ruledtabular}
\end{table}
For the pristine ZGPNRs, each inner phosphorus atom is covalently bonded with three adjacent P atoms, and each edge P atom has two bonded neighbors and one dangling bond. To explore the edge effect, in addition to the pristine ribbon, these edge P atoms were passivated using H, F, Cl, OH, O or S atoms. As an example, FIG.\ref{fig1}(c) demonstrates the snapshots of 2L-ZGPNRs with different edge passivation. Bond 1 (3) and bond 2 (4) are the nearest and next nearest P-P bonds to the edge. Bond 5 (6) is connecting to the edge passivating species. The bond lengths and angles of the ZGPNRs with different edge configurations are reported in Table \ref{tab:table2}. The bond length and angle of monolayer GP are also listed in Table \ref{tab:table2} as a reference. It is found that the length of bonds 1-4 are similar to that of monolayer GP except for the pristine one in which bond 1(3) is largely contracted to 2.15 Å. This reduced bond length in the pristine ribbons is resulted from the reconstruction of the dangling bond of the edge P atoms. The rest P-P bonds inside of the ribbons (away from the edge) has negligible change compared to that in 2D monolayer, regardless the edge passivation species. It is also noticed in Table \ref{tab:table2} that bond 2 is similar to bond 4, and bond 5 is close to bond 6, although the two edges of the ribbon structure are not completely symmetric. In addition, it was found that a larger edge chemical species yields a longer P-edge bond length b5(b6). All of the passivation except for H results in larger bond angles at the edge of the ZGPNRs compared with the original angles in the 2D GP. In general, different passivation atoms lead to slight varied distortion in the bond angles, and  $\theta_{1}$ ranged from 91.72$\degree$ to 113.56$\degree$,  $\theta_{2}$ between 99.28$\degree$ \~{} 115.47$\degree$. These observed results of geometry structures in the ZGPNRs are similar to those reported in the zigzag black phosphorene nanoribbons (ZBPNRs)\cite{RN27}.

\begin{figure}[htb]
	\includegraphics[width=8cm]{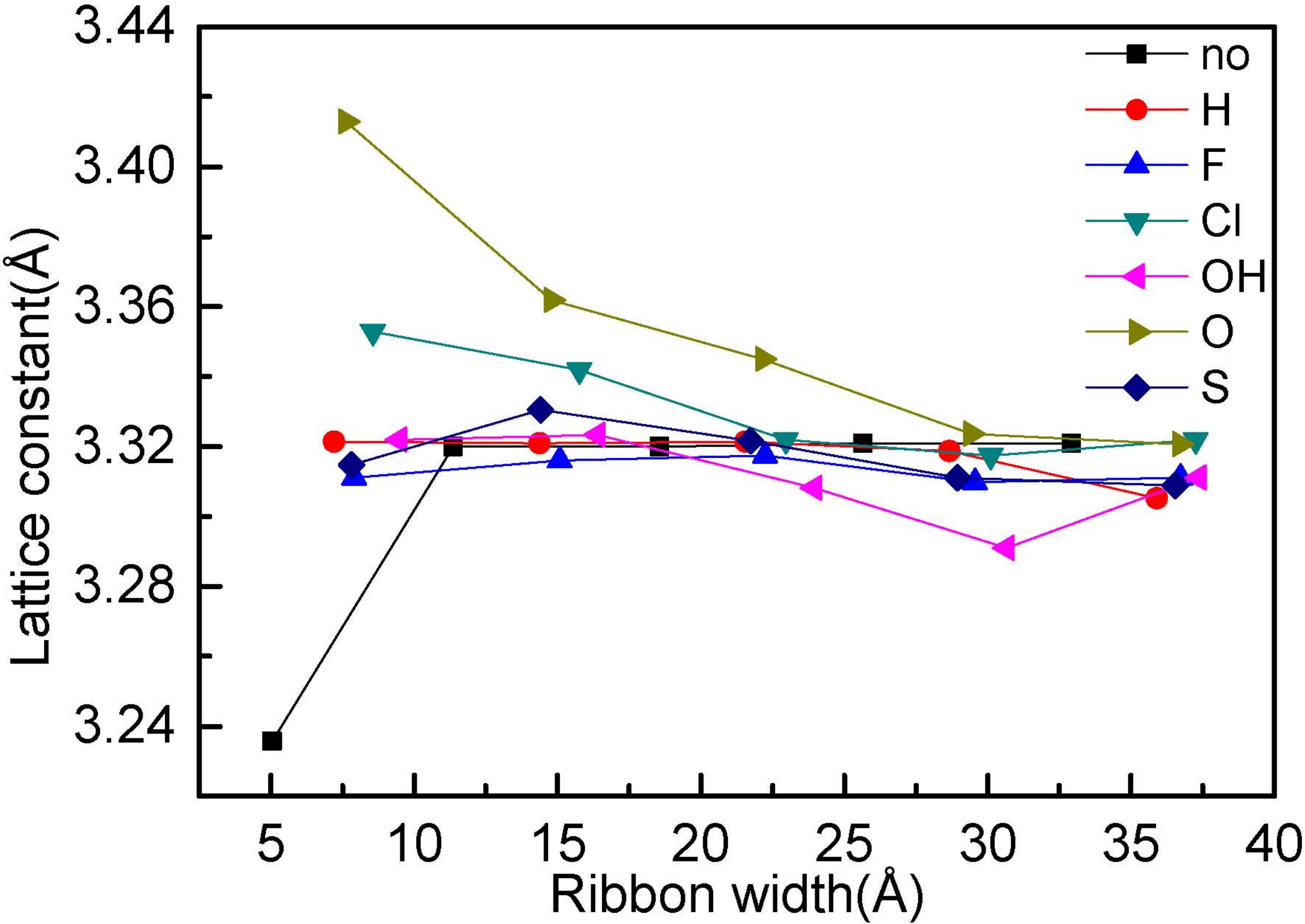}
	\caption{\label{fig2} Lattice constants of the ZGPNRs with different ribbon widths and edge functionalization groups.}
\end{figure}

\begin{table}
	\caption{\label{tab:table2}The bond lengths and angles of 2D monolayer green phosphorene and 2L-ZGPNRs with various edge passivation. }
	\begin{ruledtabular}
		\begin{tabular}{ccccccccc}
			Syetem &\tabincell{c}{b1\\(Å)} &\tabincell{c}{b2\\(Å)}&\tabincell{c}{b3\\(Å)}&\tabincell{c}{b4\\(Å)}&\tabincell{c}{b5\\(Å)}&\tabincell{c}{b6\\(Å)}&\tabincell{c}{$\theta_{1}$\\$(\degree)$}&\tabincell{c}{$\theta_{2}$\\$(\degree)$}\\
			\hline
			2D& 2.26 & 2.27 & -- & -- & -- & -- & 93.53 &102.93\\
			pristine & 2.15 & 2.26 & 2.15 & 2.26 & -- & -- & -- &--\\
			H &2.26 & 2.25 & 2.23 & 2.24 & 1.44 & 1.44 & 91.27 &99.28\\
			F &2.28& 2.26 & 2.23 & 2.25 & 1.63 & 1.64 & 97.35 &105.89\\
			Cl &2.28 &2.26&2.24 &2.24 & 2.08 & 2.07& 97.85 &107.16\\
			OH & 2.28 &2.25&2.23 & 2.25& 1.66 & 1.68 & 99.84 &105.81\\
			O & 2.27 & 2.25& 2.28& 2.26 & 1.49 & 1.50& 113.56 &115.47\\
			S &2.27 &2.24& 2.27 &2.24 & 1.98& 2.07 & 109.70 &101.20\\
		\end{tabular}
	\end{ruledtabular}
\end{table}
The relaxed lattice constants of the ZGPNRs with different ribbon widths and edge atoms were presented in FIG.\ref{fig2}. The relaxed lattice constant was obtained by scaling the lattice along the zigzag direction to reach energy minimization. As shown in FIG.\ref{fig2}, the lattice constants of the ZGPNRs depend on the ribbon widths and the specific edge chemical groups. For the cases of the H, F, or S edges, the lattice constants remain almost unchanged at 3.32 Å regardless the ribbon width, while the lattice constant with O (Cl) edge reduces with increasing ribbon widths from 3.41 Å (3.35 Å) to 3.32 Å. The variation of the lattice constants in the pristine ZGPNR is negligible except for the narrowest 1L-ZGPNR. It is 3.24 Å in the 1L-ZGPNR which is significantly smaller than that in other ribbons. Similarly, the bond angle $\alpha$ (denoted in FIG.\ref{fig2}) in 1L- ZGPNR was 97.75$\degree$ which is smaller than $\alpha = 101.17\degree$ in the 2L-ZGPNR as well. It’s also noticed that the 1L- ZGPNR owns a more puckered structure than others.

\begin{figure*}[htb]
	\includegraphics[width=16cm]{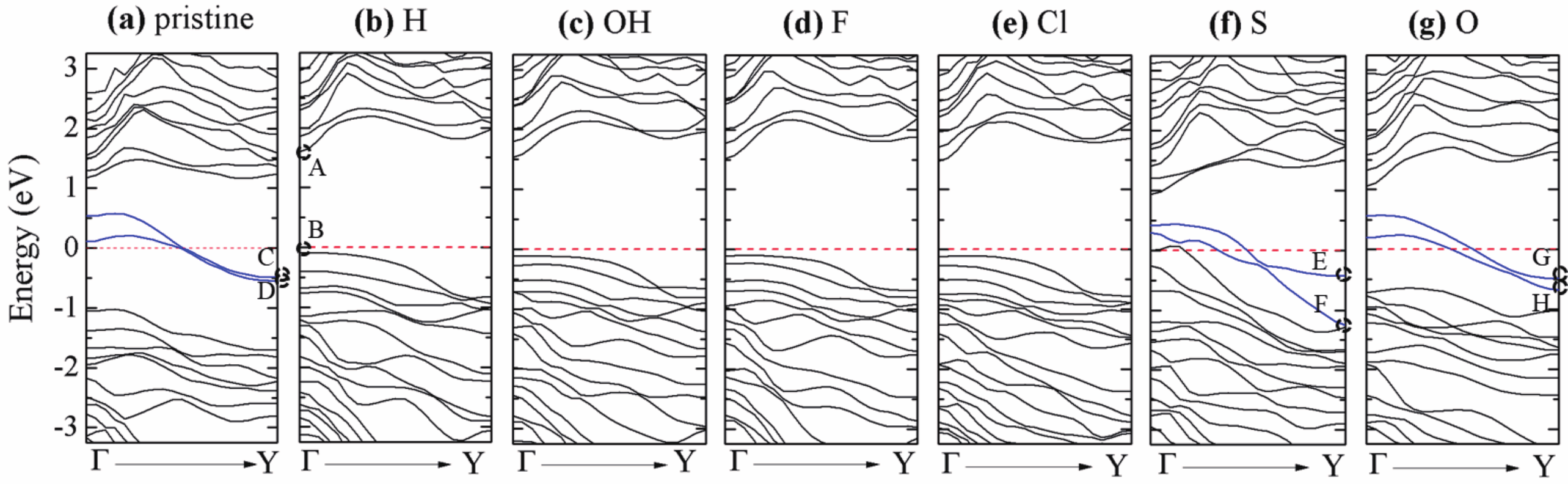}
	\caption{\label{fig3}The band structures of 2L-ZGPNRs (i.e. ribbon width 14.56 Å) with different edge configuration. The Fermi level is aligned at zero. The states brought in by the edge P, O, and S atoms crosing the Fermi level are indicated in blue lines.}
\end{figure*}

\begin{figure*}
	\includegraphics[width=16cm]{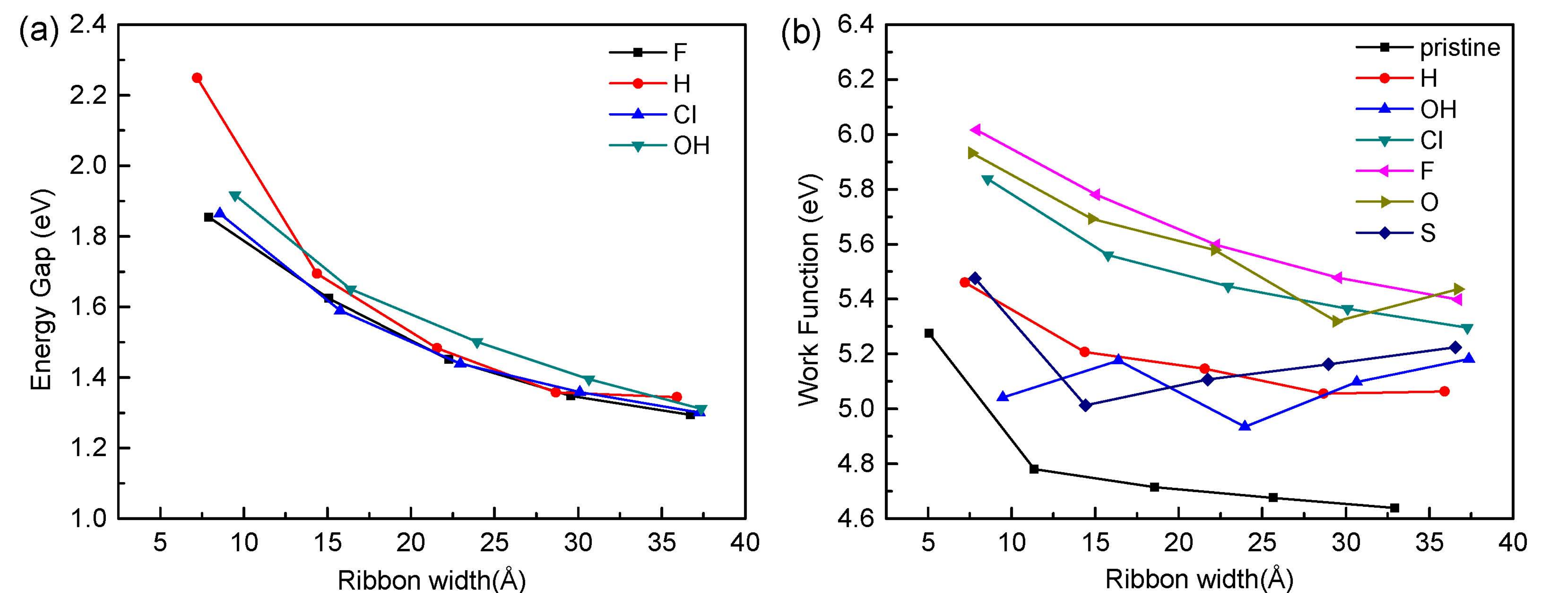}
	\caption{\label{fig4}(a) The band gap and (b) work function of the ZGPNRs with different edge functionalization were plotted as a function of ribbon width.}
\end{figure*}

{\it Electronic properties of the ZGPNRs.} The electronic band structures of the ZGPNRs with the seven different edge functionalization groups and widths were calculated to understand the edge effects. As an example, FIG.\ref{fig3} presents the band structures of the 2L-ZGPNRs. It is clear in FIG.\ref{fig3}, some ribbons possess energy states crossing the Fermi level (i.e. pristine and O, S edges), while others have finite band gaps. (i.e. the H, OH, F, and Cl edges). According to this fact, the edge chemical configurations can be classified into two distinct groups: Group \uppercase\expandafter{\romannumeral1} edges including the H, OH, F, and Cl render ribbons semiconducting, and Group \uppercase\expandafter{\romannumeral2} edges involving the pristine, O, S demonstrate metallic behavior in the ribbons. A similar results were also found in the ZBPNRs\cite{RN27}. It is clear in FIG.\ref{fig3}, for all Group \uppercase\expandafter{\romannumeral1} edges, the ZGPNRs have a direct band gap with both the CBM and VBM located at the $\Gamma$ point. The band gap of the ZGPNRs with the Group \uppercase\expandafter{\romannumeral1} edges was plotted in FIG.\ref{fig4}(a) where the band gap is tunable from 1.29 eV to 2.25 eV with the sizes and edge. The band gap reduces rapidly with the increasing ribbon widths due to the effect of quantum confinement. This direct band gap with tunability makes the ZGPNRs a promising material in applications of optoelectronic devices.

The work function of all studied ribbons was calculated and presented in FIG.\ref{fig4}(b). The work function is defined as the energy difference between vacuum and the Fermi level. As shown in FIG.\ref{fig4}(b), the work function was in the range of 4.64 \~{} 6.02 eV for the ZGPNRs with the pristine ribbons having the lowest values. The calculated work function for the 2D monolayer of GP was 5.25 eV\cite{RN17}. Since a controlled work function helps to promote the efficiency of charge transportation and photovoltaic conversion, the ZGPNRs are expect to have potential applications in photoelectro-catalyst.

\begin{figure}
	\includegraphics[width=8cm]{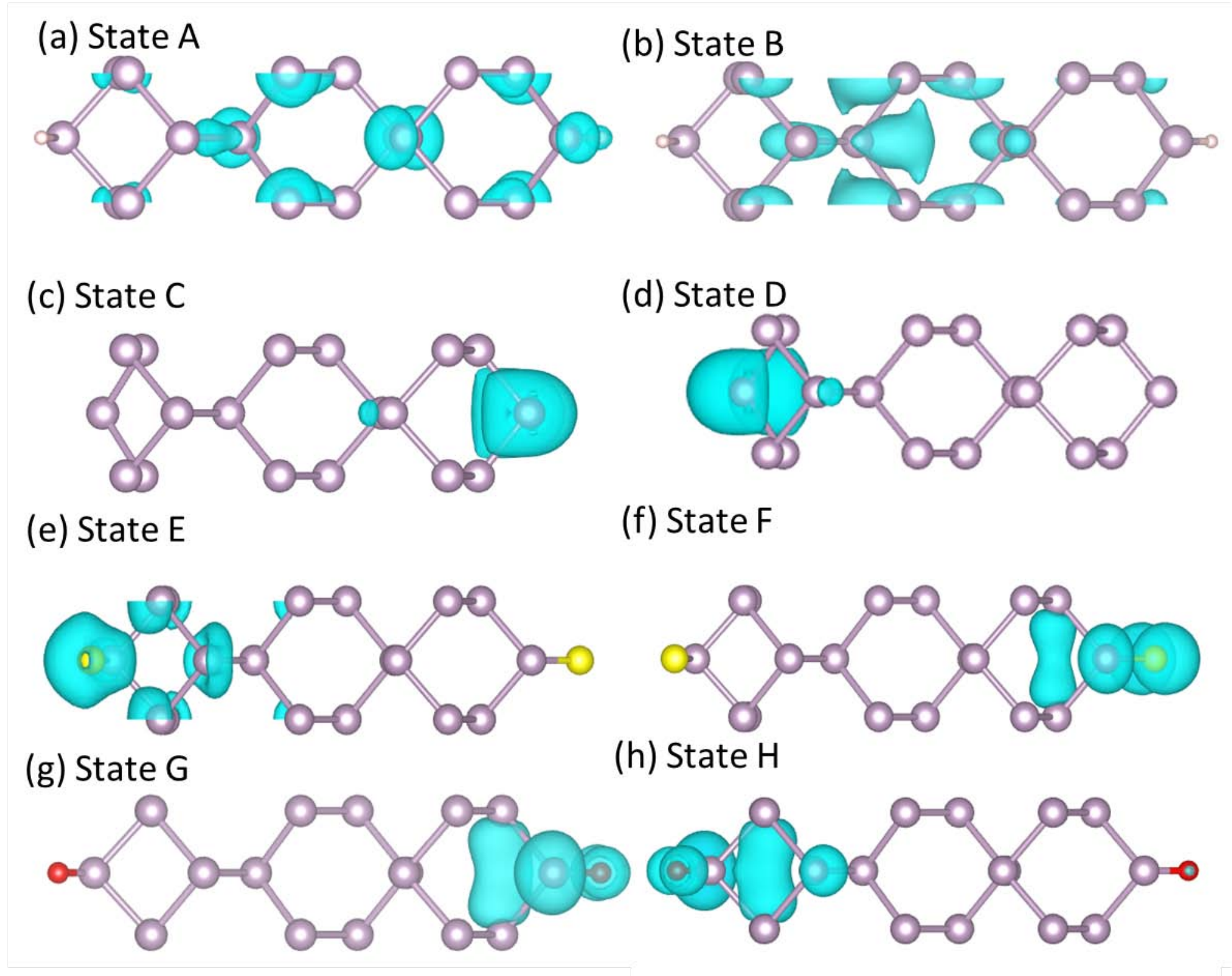}
	\caption{\label{fig5} The electron density contour plots of the states A-H (denoted in Fig.3) near the Fermi level for the 2L-ZGPNRs. The white, gray, yellow and red balls represent H, P, S, O atoms, respectively. The isosurface value is 0.005 $e/Bohr^{3}$.}
\end{figure}
To further understand the reason that the nanoribbons show either metallic or semiconducting properties with different edge configurations, a detailed analysis of the energy states near the Fermi level (denoted as states A - H in FIG.\ref{fig3}) was carried out and their electron density contour plots were presented in FIG.\ref{fig5}. States A and B presented in FIG.\ref{fig5}(a) and 5(b) are for the CBM and VBM in H-passivated 2L-ZGPNR, respectively. It is clear that the electron could of states A and B are primarily contributed by the inner P atoms and the edge P and H atoms has negligible contribution. Similarly, the CBM and VBM of the ZGPNRs with other edge passivation in Group \uppercase\expandafter{\romannumeral1} (i.e. OH, F, and Cl)are also contributed by the inner non-edge P atoms.
To explore the energy states crossing the Fermi level in the Group \uppercase\expandafter{\romannumeral2} edges, the electron density contour plots of states C-H denoted in FIG.\ref{fig3} are presented in FIG.\ref{fig5}(c) - (h) . It is clear that all the electron density of all these states are located at the edge atoms including edge P and passivation species. This fact was also observed previously in the ZBPNRs\cite{RN27}. However, there is an apparent difference in these edge states between the ZBPNRs and ZGPNRs. For the ZBPNRs, the edge states crossing the Fermi level are degenerate with the electron density located symmetrically at both edges, while for the ZGPNRs, the degeneracy of the edge states is released to have the electron cloud located at one edge due to the asymmetric structure of the ribbons.

\noindent
\underline{\it Conclusion}
Using first-principles DFT calculations, we explored the structural and electronic properties of the ZGPNRs with pristine or edges passivated using H, OH, F, Cl, O, or S edges for ribbon width up to 3.7 nm. Different cutting sequence of 1D ribbon from 2D sheet was investigated and the energies of the obtained ribbons were compared to select the most energetically favorable ones. It was found that the electronic properties of the ZGPNRs are strongly associated with the edge configurations. The ZGPNRs show either semiconducting or metallic behavior with different edges. The ZGPNRs with H, OH, F, or Cl edge are semiconductor and have a direct band gap. The CBM and VBM of the ZGPNRs with this edge Group are primarily located at the inner P atoms. The pristine ZGPNRs and the ribbons with O, or S edge show metallic behavior, in which the edge atoms bring in electronic states crossing the Fermi level.  In addition, it was found that the band gap and work function are sensitively tunable by varying the ribbon width and edge atoms.

\noindent
\underline{\it Acknowledgments}
This work was supported by NSFC (Nos. 11774033 and 11974049) and Beijing Natural Science Foundation (No.1192011).
T. M. thanks CAEP for partial financial support. Computational resources at Arizona State University Agave Cluster and Beijing Computational Science Research Center are acknowledged. Dr. Guang Yang is acknowledged for helpful discussions and review of the manuscript.

\bibliography{Reference}

\end{document}